\pdfoutput=1

\documentclass[aps,prl,twocolumn,showpacs,preprintnumbers]{revtex4}%
\usepackage{graphicx}
\usepackage{epsfig}
\usepackage{color}
\usepackage{xspace}
\usepackage{amssymb,amsmath}%
\usepackage{amsmath}%
\usepackage{amsfonts}%
\usepackage{amssymb}
\providecommand{\U}[1]{\protect\rule{.1in}{.1in}}

\definecolor{Blue}{rgb}{0.3,0.3,0.9}
\begin{document}
\title{Tailored photon-pair generation in optical fibers }
\author{Offir Cohen}
\email{o.cohen1@physics.ox.ac.uk}
\affiliation{Clarendon Laboratory, University of Oxford, Parks Road, Oxford, OX1 3PU, UK}
\author{Jeff S. Lundeen}
\affiliation{Clarendon Laboratory, University of Oxford, Parks Road, Oxford, OX1 3PU, UK}
\author{Brian J. Smith}
\affiliation{Clarendon Laboratory, University of Oxford, Parks Road, Oxford, OX1 3PU, UK}
\author{Graciana Puentes}
\affiliation{Clarendon Laboratory, University of Oxford, Parks Road, Oxford, OX1 3PU, UK}
\author{Peter J. Mosley}
\affiliation{Clarendon Laboratory, University of Oxford, Parks Road, Oxford, OX1 3PU, UK}
\author{Ian A. Walmsley}
\affiliation{Clarendon Laboratory, University of Oxford, Parks Road, Oxford, OX1 3PU, UK}
\date{\today }

\begin{abstract}
We experimentally control the spectral structure of photon pairs created via spontaneous four-wave mixing in microstructured fibers. By fabricating fibers with designed dispersion, one can manipulate the photons' wavelengths, joint spectrum, and, thus, entanglement. As an example, we produce photon-pairs with no spectral correlations, allowing direct heralding of single photons in pure-state wave packets without filtering. We achieve an experimental purity of $85.9\pm1.6\%$, while theoretical analysis and preliminary tests suggest $94.5\%$ purity is possible with a much longer fiber.
\end{abstract}

\maketitle

Optical quantum technologies such as photonic quantum computing \cite{KLM}, quantum cryptography \cite{Ekert:92}, and quantum metrology \cite{giovannetti:04} take advantage of entanglement between photons. Conventionally these technologies are based on qubit encoding, which uses a discrete variable (e.g. polarization). Entanglement can also exist between continuous variables (CVs), such as the momentum, time, position, or frequency of photons. Here we focus on frequency entanglement, which can be used to make timing measurements \cite{harris:07} and clock synchronization \cite{giovannetti:01, kuzucu:05} \ more precise, and cancel the detrimental effect of dispersion in interferometers \cite{Franson:92}. The alternative to qubit encoding, CV encoding, is a particularly powerful tool that increases the information capacity of each photon. This capacity can, for example, improve the security of quantum cryptography \cite{zhang:08} and simplify quantum computing algorithms \cite{lanyon:08}. All of these applications require different degrees and types of entanglement, and therefore, it is important to understand how to controllably create and manipulate these qualities \cite{Grice:98, Law:00, kuzucu:05, valencia:07}.

Equally important, although it has received less attention, is the difficult task of eliminating entanglement entirely, which can play a deleterious role in the production of heralded pure-state single photons. In photonic quantum computing, photon state impurity increases the error rate of logic gates by limiting the visibility of the Hong-Ou-Mandel (HOM) \cite{HOM_interferometer} interference driving these gates \cite{KLM}. Processes such as spontaneous parametric downconversion (SPDC) and spontaneous four-wave mixing (SFWM) produce photons with a small probability, but always in pairs. In \textquotedblleft heralding,\textquotedblright the detection of a photon in one beam from these sources indicates the presence of its twin and, thus, projects the other beam into a single photon state. Unfortunately, the frequencies and tranverse momenta of photon pairs produced in SPDC or SFWM are typically entangled, arising from correlations due to energy and momentum conservation constraints \cite{Law:00, Lvovsky:07}. In this case, heralding projects the other photon into an impure (i.e. ``mixed") quantum state \cite{Uren_PDC_pure}. Although this entanglement can be removed (asymptotically) by spectral and spatial filtering, doing so seriously degrades source performance in terms of production rate and heralding efficiency \cite{mosley:133601}. Generating and eliminating entanglement in photon-pair production can be seen as opposite sides of the same goal: the control of the emitted joint frequency-momentum quantum state. Here we present evidence that through the flexibility of SFWM in fiber one can reach both objectives.

In SPDC, several approaches have been developed to control, directly at the point of production, the frequency state of the emitted photon pairs. However with bulk nonlinear materials, one is limited to two control parameters: pumping geometry \cite{kuzucu:05, valencia:07} and natural material dispersion \cite{mosley:133601}. This limits our ability to tailor photon-pair frequency states. Moreover, there has been little success controlling the spatial state of the emitted photons from bulk materials, resulting in poor coupling to waveguides such as fibers.

Optical fiber sources based on SFWM are now one of the brightest photon-pair sources even when filtered \cite{Li_optical_fibre_entangled_photons} and offer significant advantages over bulk material SPDC. The photons, generated in single spatial modes, are ideal for mating with waveguides in integrated optical circuits, a promising platform for scalable photonic quantum computing \cite{Walmsley:05}. Only recently has the full potential for control of the joint frequency state of photon pairs produced by SFWM in optical fiber been brought to light \cite{Garay-Palmett:07}, and it has yet to be fully exploited experimentally \cite{Li:08}. A crucial difference from SPDC is that one can fabricate a photonic crystal fiber (PCF) to satisfy a wide range of dispersion requirements \cite{Niel} for SFWM. This new control parameter more than compensates for loss of pump geometry control that is crucial for spectral state engineering in bulk materials \cite{kuzucu:05, valencia:07}. Thus, SFWM in optical fiber allows expanded control of the spectral structure of the produced photon-pair states compared to SPDC. 

SFWM can be described as the virtual absorption of two pump ($p$) photons followed by the emission of a photon pair (the signal, $s$, and idler, $i$). The pump photons may originate from two distinct pulses or frequencies, an additional control parameter over SPDC, enabling further possibilities for state tailoring \cite{Garay-Palmett:07}. The general form of the photon-pair frequency state can be expressed as
\begin{equation} 
|\psi\rangle=\int\int d\omega_{s}\,d\omega_{i}\,f(\omega_{s},\omega
_{i})|\omega_{s}\rangle|\omega_{i}\rangle,\label{eq_general_photon_pair_state}
\end{equation}
where $|\omega_{s}\rangle|\omega_{i}\rangle$ is a photon-pair state with signal (idler) frequency $\omega_{s}$ ($\omega_{i}$), and $f(\omega_{s},\omega_{i})$ is the joint-spectral-amplitude. In this Letter, we focus on the case where both pump photons originate from the same pump pulse. Energy conservation gives rise to the dependence of $f(\omega_{s},\omega_{i})$ on the pump spectral amplitude and is assumed throughout, while momentum conservation centers the joint spectral amplitude at zero wave-vector mismatch, i.e. $\Delta k=0$, with a width inversely proportional to the length of the fiber $L$ \cite{Garay-Palmett:07}.
Here we consider a birefringent fiber \cite{Stolen:81} where the pump is polarized orthogonally to the signal and idler, in which the wave-vector mismatch, $\Delta k$ is
\begin{equation}
\Delta k=2k_{p}+\frac{2}{3}\gamma P_{p}+2\Delta n\frac{\omega_{p}}{c}-k_{s}-k_{i}.
\label{eq_phasematching}
\end{equation}
Here $k_{j}$, $j=(p,s,i)$, is the dispersion-dependent wave vector for the pump, signal and idler respectively, $P_{p}$ is the peak pump power, $\gamma$ is the fiber nonlinear coefficient, and $\Delta n$ is the birefringence. Fiber dispersion plays a central role in $\Delta k$, and thus in the joint spectral properties of the signal and idler photons created. In PCFs, the dispersion varies rapidly with frequency, allowing one to produce completely uncorrelated or highly entangled photon pairs within a small range of pump frequencies. Fiber birefringence gives an additional element of control over $\Delta k$ that has yet to be exploited experimentally, and permits one to choose the central frequencies of the photon pair states. Indeed, a key advantage of birefringent PCF fiber is that the joint spectral amplitude is completely tunable through the fiber and pump configurations. This allows the signal and idler wavelengths to be tuned within the sensitivity range of silicon-based photon-counting detectors while remaining far from the pump wavelength to eliminate contamination by Raman background, hampering several previous experiments.

\begin{figure}[h]
\centerline{\epsfig{file=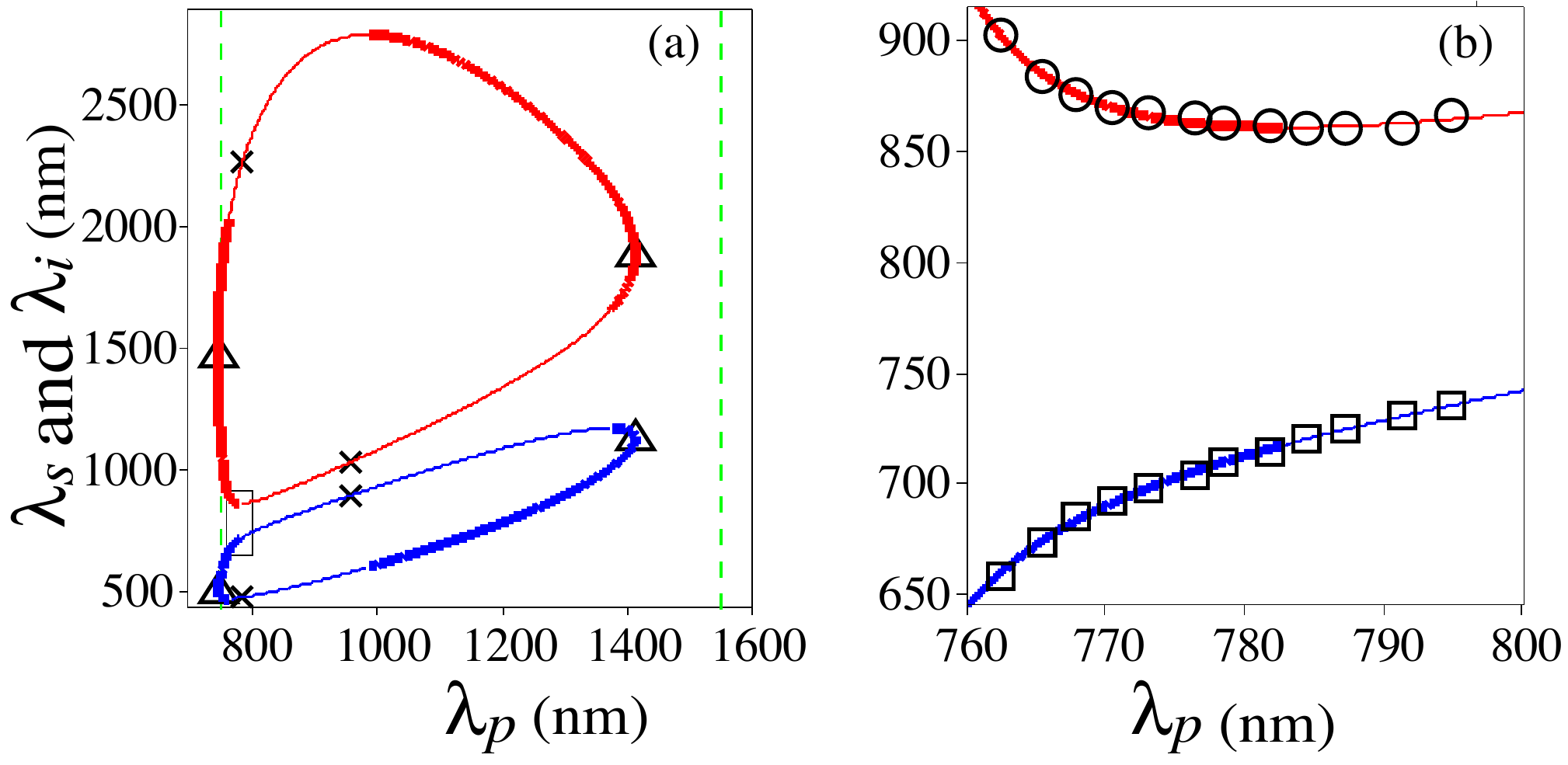, scale=0.4} }
\caption{(a) Theoretical phasematching curve (i.e. wavelength pairs where $\Delta k=0$) (solid lines) for orthogonally polarized SFWM in our fiber. (b) Magnified view of boxed region in (a). Curve is fit to experimentally measured signal (bottom $\square$) and idler (top $\bigcirc$) central wavelengths as a function of pump wavelength. Vertical dashed lines indicate the zero GVD wavelengths. Points where frequency correlated ({$\times $}) and anticorrelated states ({$\bigtriangleup$}) can be created are marked. Thin sections of the curve indicate where factorable states are possible.} 
\label{fig_spectral_curve}
\end{figure}

To demonstrate the potential of these source engineering techniques we set out to create photon pairs in an unentangled spectral state with a joint amplitude that is factorable, i.e. $f(\omega_{s},\omega_{i})=f_{s}(\omega_{s}) f_{i}(\omega_{i})$. In this situation, photon pairs are generated in only two field modes, as required for heralding pure-state single photons. To achieve this, one first fabricates or chooses a fiber in which, atypically, the group velocity of one of the generated photons is equal to that of the pump \cite{Garay-Palmett:07}. This puts tight requirements on the fiber dispersion, particularly if one wishes to choose the center wavelengths of the pump $\lambda_{p0}$ and generated photons at which this condition occurs. A signature of this group-velocity condition is that if the pump center wavelength is varied around $\lambda_{p0}$, the center wavelength of one of the generated photons remains unchanged. If the pump is at $\lambda_{p0}$ and has an appropriate bandwidth (see Ref. \cite{Garay-Palmett:07}), an unentangled spectral state will be generated.

Predicting $\lambda_{p0}$ requires precise knowledge of the fiber dispersion. Accurate birefringence and dispersion data for commercially manufactured fibers are not published, making 
direct measurement of the fiber properties necessary. Based on preliminary modeling, we chose a $40$\thinspace cm long PCF from Crystal-Fibre (model NL-1.8-750), quoted to have zero group-velocity dispersion (GVD) at $750$ and $1110$ nm, and $\gamma=99$ $[W\,km]^{-1}$ at $780$ nm. We pumped along its fast axis with a mode-locked Ti:Sapphire laser ($50$ fs full width half maximum (FWHM), $76$ MHz repetition rate). The laser bandwidth was reduced with a tuneable filter to $\Delta\lambda_{p}=4$ nm, which applies a square window function of width $\Delta\lambda_{p}$. This filter is used throughout the paper. The average power of the filtered pump was $1$\thinspace mW. The signal and idler wavelengths were recorded as the pump was tuned from $765$\thinspace nm to $795$\thinspace nm (Fig. \ref{fig_spectral_curve} (b)). This maps out the ``phasematching curve" from which the fiber dispersion and birefringence can be determined. 
Modeling the fiber as having a step index-profile \cite{wong:05} and fitting to this data yielded a core-diameter and air-filling-fraction of $1.7507$ $\mu$m and $51.1\%$ for the fast axis, and $1.7488\mu$m and $50.5\%$ for the slow axis. This difference implies a birefringence of $\Delta n\approx1.5\times10^{-5}$ at $785$\thinspace nm. From these fiber parameters, and with $P_{p}=0$, we predict a phasematching curve (Fig. \ref{fig_spectral_curve}(a)) that allows entanglement types ranging from frequency correlated to anti-correlated, all within the frequency range of a Ti:Sapphire laser pump. The experimental phasematching data shows that for $\lambda_{p} \approx785\text{ nm}$, the idler wavelength is, to first order, independent of the pump. At this point, signal and pump have the same group velocity. Although the model-based theoretical fit 
is quite accurate (less than $1\,$nm discrepancy), it predicts factorability at $\lambda_{p0}=783\text{ nm}$.
Assuming a Gaussian pump laser spectrum with a bandwidth of $20$ nm FWHM centered on $\lambda_{p}=783\text{ nm}$ that is subsequently filtered to $\Delta\lambda_{p}=8$ nm and using our fiber model we calculate the theoretical joint spectral amplitude $f(\omega_{s},\omega_{i})$. The numerical Schmidt decomposition of this result \cite{Uren_PDC_pure} predicts the purity of the heralded photons to be $86\%$.

\begin{figure}[h]
\centerline{\epsfig{file=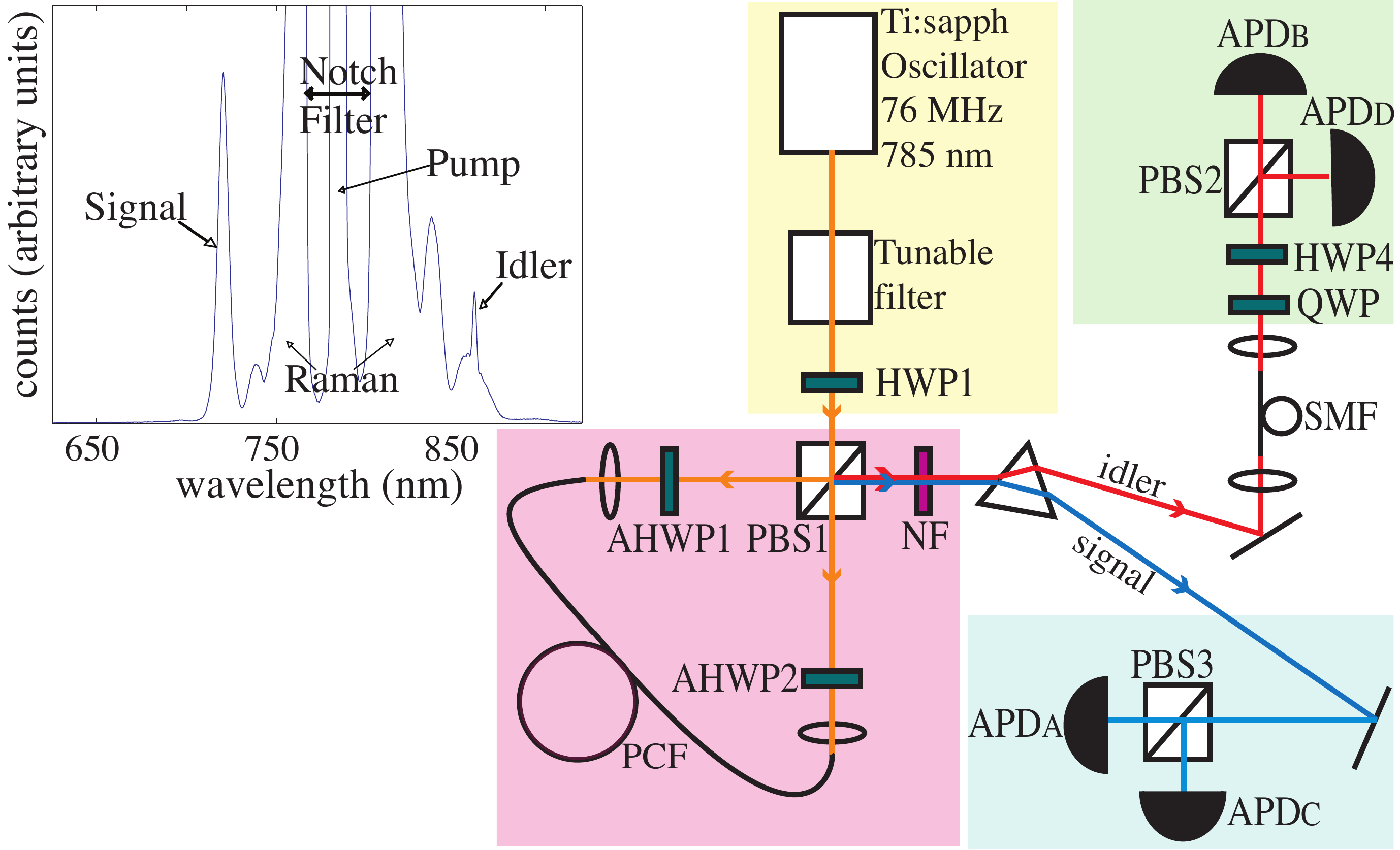, scale=0.3}}
\caption{Fiber spectrum (inset) and polarization HOM experimental setup. Although some Raman background remains at the idler wavelength, lack of background at the signal ensures coincidence detection events most likely originate from SFWM.}
\label{fig_experiment}
\end{figure}

To experimentally test the purity of a single photon state one measures the HOM\ interference between two identical copies of that state \cite{HOM_interferometer, mosley:133601}. We use a polarization analogue of the standard HOM\ interferometer: Two heralded photons, one polarized horizontally, H, and the other vertically, V, originating from independent sources and traveling the same path are rotated to $+45^{\circ}$ and $-45^{\circ}$, respectively. They are then split at a polarizing beam splitter
(PBS). The degree of suppression of the coincidence rate across the PBS output ports characterizes the purity and distinguishability of the incident photons.

We created two independent photon-pair sources with two pump pulses counter propagating through a single PCF in a Sagnac-loop configuration \cite{Rarity_SagnacLoop}, as shown in Fig. \ref{fig_experiment}. The tuneable filter reduced the pump bandwidth to $\Delta\lambda_{p}=8\,\text{nm,}$ centered on $\lambda_{p}=785\text{ nm}$ with $1.4$\thinspace mW of average power. A half waveplate (HWP1) and PBS split the pump into clockwise and counterclockwise pulses traveling through the PCF.
Achromatic HWPs (AHWP1 and AHWP2) oriented the polarizations of both pump pulses with the fast axis of the fiber and ensured the spent pump exited PBS1 back towards the laser. The photon pairs from the clockwise (H) and counterclockwise (V) directions exit from the same PBS1 port. Any remaining pump was removed with a notch filter (NF). A prism separated the signal and idler photons ($\lambda_{s(i)}=720 (860)\,\text{nm}$, $\Delta\lambda_{s(i)}=$ $3.4 (0.84)\,\text{nm}$ standard deviation) for both H and V. Occasionally, one H and one V photon pair were created simultaneously. This case was indicated by a silicon avalanche photodiode (APD) detection of two signal photons (APD{\small A} and APD{\small C}) behind PBS3. This heralded the existence of two idler photons, the subjects of our purity measurement. To ensure good spatial overlap the idler photons were passed through a single-mode fiber (SMF), which made their polarization elliptical. A quarter waveplate (QWP) partially restores linear polarization, leaving them with a residual ellipticity $\tan\chi$ (the ratio between the major and minor axes of the elliptical polarization). The polarization HOM interference was implemented by HWP4 and PBS2 with outcomes detected at APD{\small B} and
APD{\small D}.

The four-fold coincidence (detection of two signal and two idler photons) probability as a function of the polarization rotation $\theta$ induced by HWP4 is
\begin{equation}
P_{4}(\theta)=\frac{1}{2}\left[  \left(  1-p\right)  +\left(  1+p\right)
\cos^{2}(2\chi)\cos^{2}(2\theta)\right]  ,
\label{eq_four_fold}%
\end{equation}
\noindent where $p=Tr[\rho_{V}\rho_{H}]$ depends on both the indistinguishability and purity of the photons, and $\rho_{H(V)}$ is the density matrix of the heralded horizontal (vertical) photon. In the case
$\rho_{H}=\rho_{V}$, i.\thinspace e. the photons are identical, $p$ is the purity of the photons. Thus, from a fit of $P_{4}(\theta)$ we can find a lower bound on the purity.

\begin{figure}[h]
\centerline{ \epsfig{file=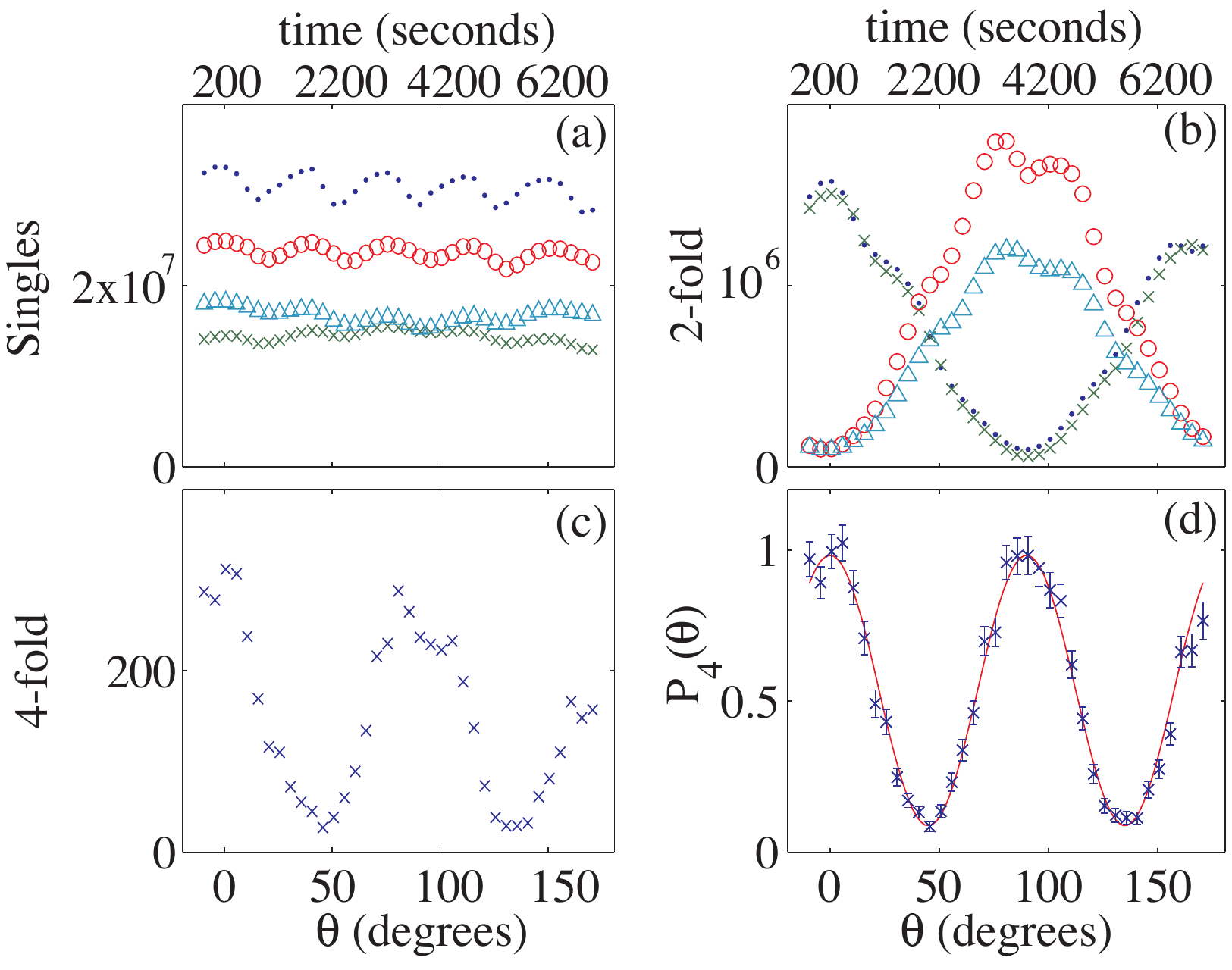, scale=0.5} }
\caption{Measured counts in $100\,$s for the polarization HOM experiment. (a) Singles counts of APDs
A ($\cdot$), B ($\times$), C ($\circ$) and D ($\bigtriangleup$). (b) Two-fold coincidence counts $R_{AB}$ ($\cdot$), $R_{CD}$ ($\times$), $R_{AD}$ ($\circ$) and $R_{BC}$ ($\bigtriangleup$). (c) Raw four-fold coincidence counts. (d) Normalized four-fold coincidence counts with theoretical fit corresponding
to purity $82.1\pm1.6\%$. Error bars represent propagated errors assuming Poissonian count statistics.}%
\label{fig_homi_short}%
\end{figure}

The raw count data as a function of $\theta$ is shown in Fig. \ref{fig_homi_short}. The singles rates oscillate due to laboratory climate control 12-minute cycling and is independent of $\theta$. Fig. \ref{fig_homi_short}(d) displays the four-fold counts normalized to account for these fluctuations according to
\begin{equation}
P_{4}(\theta)=\frac{R_{ABCD}\left(  1+\cos^{2}\left(  2\chi\right)  \cos
^{2}(2\theta)\right)  \times r\times d}{2\left[  (R_{AB}\times R_{CD}%
)+(R_{AD}\times R_{BC})\right]  },
\label{eq_four_fold_norm}%
\end{equation}
\noindent where $R_{ABCD}$ is the four-fold coincidence rate, $R_{XY}$ ($X,Y\in(A,B,C,D)$) is a two-fold coincidence rate, $d$ is the counting time for each data point, and $r$ is the laser repetition rate. Assuming identical photons, the fit of Eq.\thinspace(\ref{eq_four_fold}) to this normalized data (leaving $\chi$ free) yields an idler photon purity of $82.1\pm1.6\%$. The role of the pump time dependence and peak power in SFWM is largely untested. Varying the pump power from $0.2$ to $1.4\,$mW, we find the signal and idler central wavelengths change by only $0.5\,$nm. Moreover, we see no purity variation for power less than $1\,$mW. Above this power the observed purity decrease may be caused by higher-order processes leading to more than one photon pair being produced per pump pulse, as well as self and cross phase modulation

The measured purity in our experiment is mainly limited by the fiber length available to us. Our model predicts that the purity will approach unity as $L$ becomes large \cite{Garay-Palmett:07}; for $L$ = $100$\thinspace m the theoretical purity is $98.5\%$. Unfortunately, current commercial fibers of this length are prohibitively expensive. Alternatively, one could choose a fiber with a larger difference between pump and signal group velocities \cite{Garay-Palmett:07}.

\begin{figure}[h]
\centerline{\epsfig{file=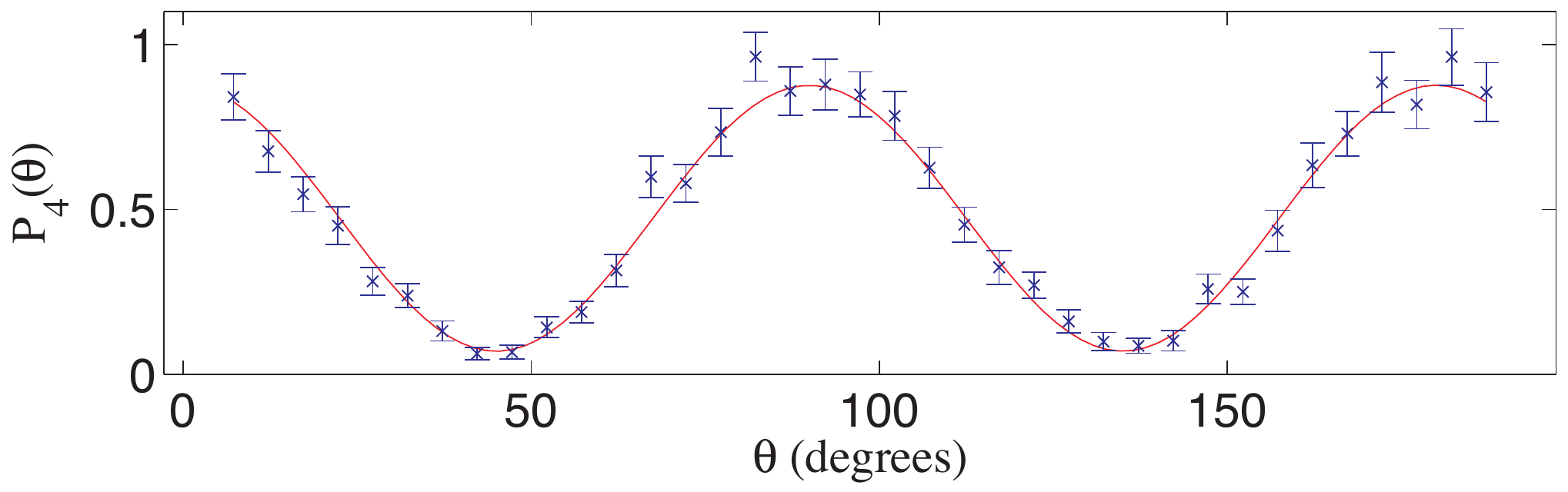, scale=0.4}}
\caption{Experimental normalized four-fold coincidences in $200\,$s with theoretical fit for $1\,$m fiber. Fit corresponds to $85.9\pm1.6\%$ purity.}%
\label{fig_homi_normalized}%
\end{figure}

To confirm this prediction, we repeated the HOM experiment using a longer fiber, $L$ $=1$\thinspace m, of the same type (see Fig. \ref{fig_homi_normalized}). The two fibers, which were obtained at different times, had factorable points $\lambda_{p0}$ that differed by $1$\thinspace nm, suggesting that uniformity might be an issue in PCF sources. Pumping with $1.4$\thinspace mW at $\lambda_{p0}=786$nm, and bandwidth $\Delta\lambda_{p}=6$\thinspace nm, we measured a purity increase of $4\%$ to $85.9\pm1.6\%$, while the purity predicted by the Schmidt decomposition method increased by $4\%$ to $90\%$. Both test fibers exhibited an experimental purity, as determined by a fit to Eq.\thinspace(\ref{eq_four_fold_norm}), of $4\%$ less than the respective theoretical predictions, suggesting a $94.5\%$ experimental purity might be possible with a 100\thinspace m fiber.

We have demonstrated the capability to tailor photon-pair states through SFWM in PCF by producing highly-pure heralded single photons without any spectral filtering, a significant source of loss in  standard sources. Eliminating this loss is critical because current photonic quantum computing circumvents loss by post selection, which is not feasible for scalable protocols. Our experiment validates recent theoretical results showing that with careful control of dispersion in a fiber, SFWM can generate a wide variety of entangled or, in our case, unentangled states \cite{Garay-Palmett:07}. PCFs provide the flexibility to produce tailored joint spectral states from the visible to telecommunications spectral regions. However, in order to facilitate concurrent use of multiple heralded photon sources, future fabrication of these fibers should address the issue of uniformity between and within fiber samples. We expect our approach to engineering the emitted optical modes to be valuable for other photon pair sources, such as atomic gases.

This work was supported by the EPSRC through the QIP IRC, by the EC under QAP
funded by IST directorate, and by the Royal Society.

\noindent Note added: Since this submission, it has come to our attention that similar results have been obtained by another research group\cite{halder}.


\begin{thebibliography}{100}
\expandafter\ifx\csname natexlab\endcsname\relax\def\natexlab#1{#1}\fi
\expandafter\ifx\csname bibnamefont\endcsname\relax
  \def\bibnamefont#1{#1}\fi
\expandafter\ifx\csname bibfnamefont\endcsname\relax
  \def\bibfnamefont#1{#1}\fi
\expandafter\ifx\csname citenamefont\endcsname\relax
  \def\citenamefont#1{#1}\fi
\expandafter\ifx\csname url\endcsname\relax
  \def\url#1{\texttt{#1}}\fi
\expandafter\ifx\csname urlprefix\endcsname\relax\def\urlprefix{URL }\fi
\providecommand{\bibinfo}[2]{#2}
\providecommand{\eprint}[2][]{\url{#2}} 


\bibitem[{\citenamefont{Knill et~al.}(2001)\citenamefont{Knill, Laflamme, and
  Milburn}}]{KLM}
\bibinfo{author}{\bibfnamefont{E.}~\bibnamefont{Knill}},
  \bibinfo{author}{\bibfnamefont{R.}~\bibnamefont{Laflamme}}, \bibnamefont{and}
  \bibinfo{author}{\bibfnamefont{G.~J.} \bibnamefont{Milburn}},
  \bibinfo{journal}{Nature} \textbf{\bibinfo{volume}{409}}, \bibinfo{pages}{46}
  (\bibinfo{year}{2001});  
\bibinfo{author}{\bibfnamefont{R.}~\bibnamefont{Raussendorf}} \bibnamefont{and}
  \bibinfo{author}{\bibfnamefont{H.~J.} \bibnamefont{Briegel}},
  \bibinfo{journal}{Phys. Rev. Lett.} \textbf{\bibinfo{volume}{86}},
  \bibinfo{pages}{5188} (\bibinfo{year}{2001}).

\bibitem[{\citenamefont{Ekert et~al.}(1992)\citenamefont{Ekert, Rarity,
  Tapster, and Massimo~Palma}}]{Ekert:92}
\bibinfo{author}{\bibfnamefont{A. K.}~\bibnamefont{Ekert}},
  \bibinfo{author}{\bibfnamefont{J. G.}~\bibnamefont{Rarity}},
  \bibinfo{author}{\bibfnamefont{P. R.}~\bibnamefont{Tapster}}, \bibnamefont{and}
  \bibinfo{author}{\bibfnamefont{G.}~\bibnamefont{Massimo~Palma}},
  \bibinfo{journal}{Phys. Rev. Lett.} \textbf{\bibinfo{volume}{69}},
  \bibinfo{pages}{1293} (\bibinfo{year}{1992}).

\bibitem[{\citenamefont{Giovannetti et~al.}(2004)\citenamefont{Giovannetti,
  Lloyd, and Maccone}}]{giovannetti:04}
\bibinfo{author}{\bibfnamefont{V.}~\bibnamefont{Giovannetti}},
  \bibinfo{author}{\bibfnamefont{S.}~\bibnamefont{Lloyd}}, \bibnamefont{and}
  \bibinfo{author}{\bibfnamefont{L.}~\bibnamefont{Maccone}},
  \bibinfo{journal}{Science} \textbf{\bibinfo{volume}{306}},
  \bibinfo{pages}{1330} (\bibinfo{year}{2004}); 
  \bibinfo{author}{\bibfnamefont{M.~W.} \bibnamefont{Mitchell}},
  \bibinfo{author}{\bibfnamefont{J.~S.} \bibnamefont{Lundeen}},
  \bibnamefont{and} \bibinfo{author}{\bibfnamefont{A.~M.}
  \bibnamefont{Steinberg}}, \bibinfo{journal}{Nature}
  \textbf{\bibinfo{volume}{429}}, \bibinfo{pages}{161} (\bibinfo{year}{2004}).

\bibitem[{\citenamefont{Harris}(2007)}]{harris:07}
\bibinfo{author}{\bibfnamefont{K.~A.} \bibnamefont{O'Donnell}}
  \bibnamefont{and} \bibinfo{author}{\bibfnamefont{A.~B.} \bibnamefont{U'Ren}},
  \bibinfo{journal}{Opt. Lett.} \textbf{\bibinfo{volume}{32}},
  \bibinfo{pages}{817} (\bibinfo{year}{2007});
\bibinfo{author}{\bibfnamefont{S. E.}~\bibnamefont{Harris}},
  \bibinfo{journal}{Phys. Rev. Lett.} \textbf{\bibinfo{volume}{98}},
  \bibinfo{pages}{63602} (\bibinfo{year}{2007});   \bibinfo{journal}{ibid.} \textbf{\bibinfo{volume}{100}},
  \bibinfo{pages}{183601} (\bibinfo{year}{2008}).

\bibitem[{\citenamefont{GIOVANNETTI et~al.}(2001)\citenamefont{GIOVANNETTI,
  LLOYD, and MACCONE}}]{giovannetti:01}
\bibinfo{author}{\bibfnamefont{V.}~\bibnamefont{Giovannetti}},
  \bibinfo{author}{\bibfnamefont{S.}~\bibnamefont{Lloyd}}, \bibnamefont{and}
  \bibinfo{author}{\bibfnamefont{L.}~\bibnamefont{Maccone}},
  \bibinfo{journal}{Nature} \textbf{\bibinfo{volume}{412}},
  \bibinfo{pages}{417} (\bibinfo{year}{2001}); 
\bibinfo{author}{\bibfnamefont{V.}~\bibnamefont{Giovannetti}},
  \bibinfo{author}{\bibfnamefont{L.}~\bibnamefont{Maccone}},
  \bibinfo{author}{\bibfnamefont{J. H.}~\bibnamefont{Shapiro}}, \bibnamefont{and}
  \bibinfo{author}{\bibfnamefont{F. N. C.}~\bibnamefont{Wong}},
  \bibinfo{journal}{Phys. Rev. Lett.} \textbf{\bibinfo{volume}{88}},
  \bibinfo{pages}{183602} (\bibinfo{year}{2002}).

\bibitem[{\citenamefont{Kuzucu et~al.}(2005)\citenamefont{Kuzucu, Fiorentino,
  Albota, Wong, and K{\"a}rtner}}]{kuzucu:05}
\bibinfo{author}{\bibfnamefont{O.}~\bibnamefont{Kuzucu et. al.}},
  \bibinfo{journal}{Phys. Rev. Lett.} \textbf{\bibinfo{volume}{94}},
  \bibinfo{pages}{83601} (\bibinfo{year}{2005}).

\bibitem[{\citenamefont{Franson}(1992)}]{Franson:92}
\bibinfo{author}{\bibfnamefont{J.~D.} \bibnamefont{Franson}},
  \bibinfo{journal}{Phys. Rev. A} \textbf{\bibinfo{volume}{45}},
  \bibinfo{pages}{3126} (\bibinfo{year}{1992}); 
\bibinfo{author}{\bibfnamefont{A.~M.} \bibnamefont{Steinberg}},
  \bibinfo{author}{\bibfnamefont{P.~G.} \bibnamefont{Kwiat}}, \bibnamefont{and}
  \bibinfo{author}{\bibfnamefont{R.~Y.} \bibnamefont{Chiao}},
  \bibinfo{journal}{Phys. Rev. Lett.} \textbf{\bibinfo{volume}{68}},
  \bibinfo{pages}{2421} (\bibinfo{year}{1992}); 
\bibinfo{author}{\bibfnamefont{M.~B.} \bibnamefont{Nasr}},
  \bibinfo{author}{\bibfnamefont{B.~E.~A.} \bibnamefont{Saleh}},
  \bibinfo{author}{\bibfnamefont{A.~V.} \bibnamefont{Sergienko}},
  \bibnamefont{and} \bibinfo{author}{\bibfnamefont{M.~C.} \bibnamefont{Teich}},
  \bibinfo{journal}{ibid.} \textbf{\bibinfo{volume}{91}},
  \bibinfo{pages}{083601} (\bibinfo{year}{2003}).
  
  
\bibitem[{\citenamefont{Zhang et~al.}(2008)\citenamefont{Zhang, Silberhorn, and
  Walmsley}}]{zhang:08}
\bibinfo{author}{\bibfnamefont{L.}~\bibnamefont{Zhang}},
  \bibinfo{author}{\bibfnamefont{C.}~\bibnamefont{Silberhorn}},
  \bibnamefont{and} \bibinfo{author}{\bibfnamefont{I.~A.}
  \bibnamefont{Walmsley}}, \bibinfo{journal}{Phys. Rev. Lett.}
  \textbf{\bibinfo{volume}{100}}, \bibinfo{eid}{110504}
  (\bibinfo{year}{2008}).

\bibitem[{\citenamefont{Lanyon et~al.}(2008)\citenamefont{Lanyon, Barbieri,
  Almeida, Jennewein, Ralph, Resch, Pryde, O'Brien, Gilchrist, and
  White}}]{lanyon:08}
\bibinfo{author}{\bibfnamefont{B.~P.} \bibnamefont{Lanyon et. al.}},
  \bibinfo{journal}{arXiv:0804.0272v1}  (\bibinfo{year}{2008}).


\bibitem[{\citenamefont{Grice et~al.}(1998)\citenamefont{Grice, Erdmann,
  Walmsley, and Branning}}]{Grice:98}
\bibinfo{author}{\bibfnamefont{W. P.}~\bibnamefont{Grice}},
  \bibinfo{author}{\bibfnamefont{R.}~\bibnamefont{Erdmann}},
  \bibinfo{author}{\bibfnamefont{I. A.}~\bibnamefont{Walmsley}}, \bibnamefont{and}
  \bibinfo{author}{\bibfnamefont{D.}~\bibnamefont{Branning}},
  \bibinfo{journal}{Phys. Rev. A} \textbf{\bibinfo{volume}{57}},
  \bibinfo{pages}{R2289} (\bibinfo{year}{1998}).

\bibitem[{\citenamefont{Law et~al.}(2000)\citenamefont{Law, Walmsley, and
  Eberly}}]{Law:00}
\bibinfo{author}{\bibfnamefont{C.~K.} \bibnamefont{Law}},
  \bibinfo{author}{\bibfnamefont{I.~A.} \bibnamefont{Walmsley}},
  \bibnamefont{and} \bibinfo{author}{\bibfnamefont{J.~H.}
  \bibnamefont{Eberly}}, \bibinfo{journal}{Phys. Rev. Lett.}
  \textbf{\bibinfo{volume}{84}}, \bibinfo{pages}{5304} (\bibinfo{year}{2000}); 
\bibinfo{author}{\bibfnamefont{W.~P.} \bibnamefont{Grice}},
  \bibinfo{author}{\bibfnamefont{A.~B.} \bibnamefont{U'Ren}},
  \bibnamefont{and} \bibinfo{author}{\bibfnamefont{I.~A.}
  \bibnamefont{Walmsley}}, \bibinfo{journal}{Phys. Rev. A}
  \textbf{\bibinfo{volume}{64}}, \bibinfo{pages}{063815}
  (\bibinfo{year}{2001}).

\bibitem[{\citenamefont{Valencia et~al.}(2007)\citenamefont{Valencia, Cer{\'e},
  Shi, Molina-Terriza, and Torres}}]{valencia:07}
\bibinfo{author}{\bibfnamefont{A.}~\bibnamefont{Valencia et. al.}},
  \bibinfo{journal}{Phys. Rev. Lett.} \textbf{\bibinfo{volume}{99}},
  \bibinfo{pages}{243601} (\bibinfo{year}{2007}).


\bibitem[{\citenamefont{Hong et~al.}(1987)\citenamefont{Hong, Ou, and
  Mandel}}]{HOM_interferometer}
\bibinfo{author}{\bibfnamefont{C.~K.} \bibnamefont{Hong}},
  \bibinfo{author}{\bibfnamefont{Z.~Y.} \bibnamefont{Ou}}, \bibnamefont{and}
  \bibinfo{author}{\bibfnamefont{L.}~\bibnamefont{Mandel}},
  \bibinfo{journal}{Phys. Rev. Lett.} \textbf{\bibinfo{volume}{59}}, \bibinfo{eid}{2044}
  (\bibinfo{year}{1987}).
  
  
  \bibitem[{\citenamefont{Lvovsky et~al.}(2007)\citenamefont{Lvovsky, Wasilewski,
  and Banaszek}}]{Lvovsky:07}
\bibinfo{author}{\bibfnamefont{A.~I.} \bibnamefont{Lvovsky}},
  \bibinfo{author}{\bibfnamefont{W.}~\bibnamefont{Wasilewski}},
  \bibnamefont{and} \bibinfo{author}{\bibfnamefont{K.}~\bibnamefont{Banaszek}},
  \bibinfo{journal}{J. Mod. Opt.} \textbf{\bibinfo{volume}{54}},
  \bibinfo{pages}{721} (\bibinfo{year}{2007}).

\bibitem[{\citenamefont{U'Ren et~al.}(2005)\citenamefont{U'Ren, Silberhorn,
  Banaszek, Walmsley, Erdmann, Grice, and Raymer}}]{Uren_PDC_pure}
\bibinfo{author}{\bibfnamefont{A.~B.} \bibnamefont{U'Ren et. al.}},
  \bibinfo{journal}{Las. Phys.} \textbf{\bibinfo{volume}{15}},
  \bibinfo{pages}{146} (\bibinfo{year}{2005}).


\bibitem[{\citenamefont{Mosley et~al.}(2008)\citenamefont{Mosley, Lundeen,
  Smith, Wasylczyk, U'Ren, Silberhorn, and Walmsley}}]{mosley:133601}
\bibinfo{author}{\bibfnamefont{P.~J.} \bibnamefont{Mosley et. al.}},
  \bibinfo{journal}{Phys. Rev. Lett.}
  \textbf{\bibinfo{volume}{100}}, \bibinfo{eid}{133601}
  (\bibinfo{year}{2008}).

\bibitem[{\citenamefont{Li et~al.}(2005)\citenamefont{Li, Voss, Sharping, and
  Kumar}}]{Li_optical_fibre_entangled_photons}
\bibinfo{author}{\bibfnamefont{X.}~\bibnamefont{Li}},
  \bibinfo{author}{\bibfnamefont{P.~L.} \bibnamefont{Voss}},
  \bibinfo{author}{\bibfnamefont{J.~E.} \bibnamefont{Sharping}},
  \bibnamefont{and} \bibinfo{author}{\bibfnamefont{P.}~\bibnamefont{Kumar}},
  \bibinfo{journal}{Phys. Rev. Lett.} \textbf{\bibinfo{volume}{94}},
  \bibinfo{pages}{053601} (\bibinfo{year}{2005}); 
\bibinfo{author}{\bibfnamefont{J.}~\bibnamefont{Fan}},
  \bibinfo{author}{\bibfnamefont{A.}~\bibnamefont{Migdall}}, \bibnamefont{and}
  \bibinfo{author}{\bibfnamefont{L.~J.} \bibnamefont{Wang}},
  \bibinfo{journal}{Opt. Lett.} \textbf{\bibinfo{volume}{30}},
  \bibinfo{pages}{3368} (\bibinfo{year}{2005}); 
\bibinfo{author}{\bibfnamefont{J.}~\bibnamefont{Fulconis et. al.}},
  \bibinfo{journal}{New J. Phys.}
  \textbf{\bibinfo{volume}{9}}, \bibinfo{eid}{276}
  (\bibinfo{year}{2007}).

\bibitem[{\citenamefont{Walmsley and Raymer}(2005)}]{Walmsley:05}
\bibinfo{author}{\bibfnamefont{I.~A.} \bibnamefont{Walmsley}} \bibnamefont{and}
  \bibinfo{author}{\bibfnamefont{M.~G.} \bibnamefont{Raymer}},
  \bibinfo{journal}{Science} \textbf{\bibinfo{volume}{307}},
  \bibinfo{pages}{1733} (\bibinfo{year}{2005}); 
\bibinfo{author}{\bibfnamefont{A.}~\bibnamefont{Politi et. al.}},
  \bibinfo{journal}{ibid.} \textbf{\bibinfo{volume}{320}},
  \bibinfo{pages}{646}
  (\bibinfo{year}{2008}).

\bibitem[{\citenamefont{Garay-Palmett et~al.}(2007)\citenamefont{Garay-Palmett,
  McGuinness, Cohen, Lundeen, Rangel-Rojo, U'ren, Raymer, McKinstrie, Radic,
  and Walmsley}}]{Garay-Palmett:07}
\bibinfo{author}{\bibfnamefont{K.}~\bibnamefont{Garay-Palmett et. al.}},
  \bibinfo{journal}{Opt. Express} \textbf{\bibinfo{volume}{15}},
  \bibinfo{pages}{14870} (\bibinfo{year}{2007}).

\bibitem[{\citenamefont{Li et~al.}(2008)\citenamefont{Li, Ma, Ou, Yang, Cui,
  and Yu}}]{Li:08}
\bibinfo{author}{\bibfnamefont{X.}~\bibnamefont{Li et. al.}},
  \bibinfo{journal}{Opt. Express} \textbf{\bibinfo{volume}{16}}, \bibinfo{pages}{32}
  (\bibinfo{year}{2008}).
  
  \bibitem[{\citenamefont{Poletti et. al.}(2008)\citenamefont{F. Poletti et. al.}}]{Niel}
\bibinfo{author}{\bibfnamefont{F.}~\bibnamefont{Poletti et. al.}},
   \bibinfo{journal}{Opt. Express} \textbf{\bibinfo{volume}{13}}, \bibinfo{pages}{3728}
  (\bibinfo{year}{2005}).

\bibitem[{\citenamefont{Stolen et~al.}(1981)\citenamefont{Stolen, Bosch, and
  Lin}}]{Stolen:81}
\bibinfo{author}{\bibfnamefont{R.~H.} \bibnamefont{Stolen}},
  \bibinfo{author}{\bibfnamefont{M.~A.} \bibnamefont{Bosch}}, \bibnamefont{and}
  \bibinfo{author}{\bibfnamefont{C.}~\bibnamefont{Lin}}, \bibinfo{journal}{Opt.
  Lett.} \textbf{\bibinfo{volume}{6}}, \bibinfo{pages}{213}
  (\bibinfo{year}{1981}).

\bibitem[{\citenamefont{Wong et~al.}(2005)\citenamefont{Wong, Chen, Ha,
  Kruhlak, Murdoch, Leonhardt, Harvey, and Joly}}]{wong:05}
\bibinfo{author}{\bibfnamefont{G.~K.} \bibnamefont{Wong et. al.}},
  \bibinfo{journal}{Opt. Express} \textbf{\bibinfo{volume}{13}},
  \bibinfo{pages}{8662} (\bibinfo{year}{2005}).

\bibitem[{\citenamefont{Fulconis
  et~al.}(2007{\natexlab{b}})\citenamefont{Fulconis, Alibart, O'Brien,
  Wadsworth, and Rarity}}]{Rarity_SagnacLoop}
\bibinfo{author}{\bibfnamefont{J.}~\bibnamefont{Fulconis et. al.}},
  \bibinfo{journal}{Phys. Rev. Lett.}
  \textbf{\bibinfo{volume}{99}}, \bibinfo{eid}{120501}
  (\bibinfo{year}{2007}); 
\bibinfo{author}{\bibfnamefont{J.}~\bibnamefont{Fan}},
  \bibinfo{author}{\bibfnamefont{M.~D.} \bibnamefont{Eisaman}},
  \bibnamefont{and} \bibinfo{author}{\bibfnamefont{A.}~\bibnamefont{Migdall}},
  \bibinfo{journal}{Phys. Rev. A}
  \textbf{\bibinfo{volume}{76}}, \bibinfo{eid}{043836}
  (\bibinfo{year}{2007}).

\bibitem[{\citenamefont{Fulconis
  et~al.}(2007{\natexlab{b}})\citenamefont{Fulconis, Alibart, O'Brien,
  Wadsworth, and Rarity}}]{halder}
\bibinfo{author}{\bibfnamefont{M.}~\bibnamefont{Halder \emph{et al.}}}
\bibinfo{journal}{Opt. Express} \textbf{\bibinfo{volume}{17}}, \bibinfo{eid}{4670}
  (\bibinfo{year}{2009}). 

\end{thebibliography}
\end{document}